\documentstyle{article}

\def\abstract#1{\vskip 7mm 
        \begin{center}{\large Abstract}\par \smallskip
                \begin{minipage}[c]{12cm}
                        \small #1
                \end{minipage}
        \end{center}
}
\def\title#1{\begin{center}{\Large\bf #1}\end{center}}
\def\author#1{\vskip 5mm \begin{center}{#1}\end{center}}
\def\address#1{\begin{center}{\it #1}\end{center}}
\makeatletter
\@ifundefined{lesssim}{}{}
\@ifundefined{gtrsim}{}{}
\def\vereq#1#2{\lower3pt\vbox{\baselineskip1.5pt \lineskip1.5pt
\ialign{$\m@th#1\hfill##\hfil$\crcr#2\crcr\sim\crcr}}}
\makeatother

\begin{document}

\title{%
  Stars on  branes~:
the view from the brane
  \smallskip \\
}
\author{%
 Nathalie Deruelle\footnote{E-mail:deruelle@iap.fr}
  }
\address{%
  D\'epartement d'Astrophysique Relativiste et de Cosmologie,\\
  UMR 8629 du Centre National de la Recherche Scientifique,\\
Observatoire de Paris, 92195 Meudon, France
}

\address{
 Institut des Hautes Etudes Scientifiques, \\ 
 91440, Bures-sur-Yvette, France}

\centerline{ submitted on October 30th to the proceedings in honour of
Prof. Tomita's retirement}

\abstract{\bigskip
  We consider  spherically symmetric matter configurations on a four dimensional  ``brane" embedded in a five dimensional $Z_2$-symmetric ``bulk". We write the junction
conditions between the interior and exterior of these ``stars", treat a couple of static examples in order to point out the differences with ordinary four dimensional Einstein
gravity, consider briefly a collapse situation and conclude with the importance of a global view including  asymptotic and regularity conditions in the bulk.
}
\bigskip

\section{Introduction}

Consider  a smooth five dimensional spacetime ${\cal V}_5$ embedded, for a better vizualisation, in a six dimensional space~; consider  a four dimensional timelike
hypersurface
$M_4$ in  ${\cal V}_5$~; cut
${\cal V}_5$ along
$M_4$~; make a copy of one of the two ``halves" of  ${\cal V}_5$ and ``paste" it along $M_4$. The resulting manifold is a $Z_2$-symmetric ``bulk", ${\cal M}_5$. The
cutting and pasting procedure turns $M_4$ into a singular ``brane" whose extrinsic curvature in ${\cal M}_5$ exhibits a discontinuity which is twice the value of the
(regular) extrinsic curvature of
$M_4$ in ${\cal V}_5$. The  Einstein tensor of ${\cal M}_5$ which contains derivatives of the extrinsic curvature therefore exhibits a
delta like singularity whose coefficient can be expressed in terms of the jump of the extrinsic curvature [1] and is interpreted, within Einstein's theory of gravity, as the sum of
the tension of the brane and the stress-energy tensor of its matter content [2]. If ${\cal V}_5$ is an Einstein space, that is if its Ricci tensor is proportional to its
metric, the four dimensional Einstein tensor $G_{\mu\nu}$ of the brane
$M_4$ can hence be written as (see [3] and, e.g., [4])
$$G_{\mu\nu}=8\pi G\, T_{\mu\nu}-{\cal E}_{\mu\nu}+{\kappa^2\over4}\left[-T_{\mu\rho}T^\rho_\nu+{1\over3}TT_{\mu\nu}+{1\over2}g_{\mu\nu}\left(T.
T-{1\over3}T^2\right)\right]\eqno(1.1)$$ 
where $G$ is Newton's constant, $\kappa$ a coupling constant related to $G$ and the (constant) scalar curvature of the bulk, where $g_{\mu\nu}$ is the
metric of $M_4$  (and $D$ its associated covariant derivative),  where $T_{\mu\nu}$ is related to the jump of the extrinsic curvature of $M_4$ and identified to the {\it
conserved} ($D_\mu T^\mu_\nu=0$) stress-energy tensor of matter in
$M_4$ (with the notation
$T\equiv T_\rho^\rho$ and $T.T\equiv T_{\rho\sigma}T^{\rho\sigma}$), and where ${\cal E}_{\mu\nu}$ is the {\it traceless} (${\cal E}=0$) projection of the bulk Weyl tensor onto
the brane. Equation (1.1) is the equation governing gravity on the brane. It differs from the standard four dimensional Einstein equation by the presence of the term quadratic in
$T_{\mu\nu}$ and the Weyl term.

In the following we shall study (1.1) when the brane is spherically symmetric and describes a ``star", either static (Section III) or collapsing (Section IV). Hence we suppose that
the brane is empty (that is, $T_{\mu\nu}=0$) outside the boundary of the star, and is filled with some fluid  inside. Such a problem has already
been studied by a few authors. In [5], Germani and Maartens consider a static configuration~; in [6], Bruni et al. investigate a collapse situation and  Govender and
Dadhich  extend the results in [7]. Numerical analyses are also underway (Turok and Wiseman, private communication). The contribution of the present paper is to show in Section
II that the junction conditions across the boundary of the star on the brane are different from the standard general relativistic ones and allow (at least in principle)  a wider
range of configurations than those considered in [5-7]. We also approach in Section V the problem of describing the geometry of the bulk and try to give a few insights on how the
condition of a smooth and asymptotically anti-de Sitter bulk severely constrains the possible star configurations on the brane.

\section{Junction conditions on the ``star" boundary}

Let us use, in the brane $M_4$, a gaussian normal coordinate system near the surface of the star, that is, write the brane metric as $ds^2=dz^2+g_{ab}(z,x^c)dx^a dx^b$, where
$z=0$ is the (spherically symmetric, timelike) surface $S$ of the star and  $x^a$ three coordinates on $S$. Introducing $k_{ab}\equiv \partial g_{ab}/\partial z$
and $l_{ab}\equiv\partial^2 g_{ab}/\partial z^2$, the Einstein tensor of $M_4$ hence decomposes as
$$G_{ab}=\,^{(3)}\!G_{ab}-{1\over2}(l_{ab}-l\,g_{ab})+{1\over2}k_{ac}k^c_b-{1\over4}kk_{ab}-{1\over8}g_{ab}(3k.k-k^2)\eqno(2.1)$$
$$G_{az}={1\over2}(\nabla_ck^c_a-\nabla_ak)\eqno(2.2)$$
$$G_{zz}=-{1\over2}\,^{(3)}\!R-{1\over8}(k.k-k^2)\eqno(2.3)$$
where $\nabla$, $\,^{(3)}\!G_{ab}$ and $\,^{(3)}\!R$ are the covariant derivative, the Einstein tensor and the scalar curvature of the three metric $g_{ab}$. 

Now the equations for gravity are (1.1), with  $T_{\mu\nu}(z,x^c)$ exhibiting a Heaviside type discontinuity across $S$. A first problem is that the right
hand side of (1.1) includes terms quadratic in $T_{\mu\nu}$ which are {\it not} defined in a distributional sense. We shall {\it suppose} however that these terms are still of
the Heaviside type. A second problem is that, from a
brane point of view, little is known about
${\cal E}_{\mu\nu}$ which can a priori exhibit any kind of discontinuity across $S$ (such as a thin layer). We shall {\it suppose} here that ${\cal E}_{\mu\nu}$ is also a Heaviside
distribution. 

With these hypotheses equation (1.1) combined with (2.1) implies that $l_{ab}$ is a Heaviside distribution, so that $k_{ab}$ and $g_{ab}$ are continuous across $S$. More
geometrically, that is in  arbitrary coordinate systems on each side of $S$, this condition translates as the familiar condition of the continuity of the induced metrics on $S$
and the continuity of the extrinsic curvatures of $S$ [1]. As for (1.1) combined with (2.2) and (2.3) they tell us that the $(za)$ and $(zz)$ components of the right hand side of
(1.1) are continuous across $S$. These conditions are redundant because of the Bianchi identity but they can be handy to consider as well since they read as conditions on
the matter in $S$ rather than on the geometry of $S$.
 For example, they show clearly that,
contrarily to what happens in standard general relativity, the ``pressure" $T_{zz}$ is not necessarily zero on $S$. Imposing $T_{zz}|_S=0$, as is done in [5-7], may be a
physically well motivated assumption but is not {\it imposed} by the equations for gravity on the brane. In the following we shall hence allow for non zero  pressures on
the boundary of the star.

\section{Static ``stars" on a brane}

Outside the star the brane is spherically symmetric and static. In Schwarzschild coordinates $(t,r,\theta,\phi)$ and setting $d\Omega^2=d\theta^2+\sin^2\!\theta\, d\phi^2$ the
metric is of the form
$$ds^2_e=-\Phi^e(r)dt^2+{dr^2\over\Psi^e(r)}+r^2d\Omega^2\,.$$ 
Similarly, inside the star where the Schwarzschild coordinates are called $(T,R,\theta,\phi$) we have
$$ds^2_i=-\Phi^i(R)dT^2+{dR^2\over\Psi^i(R)}+R^2d\Omega^2\,.$$
The surface of the star is defined by $r=r_0$ and $R=R_0$. The assumed continuity of the induced metrics and extrinsic curvatures imposes the same geometrical conditions as in
four dimensional general relativity, that is
$$r_0=R_0\eqno(3.1)$$
$$\Psi^e_0=\Psi^i_0\eqno(3.2)$$
$${\Phi^{e'}_0\over\Phi^e_0}={\Phi^{i'}_0\over\Phi^i_0}\eqno(3.3)$$
where a prime indicates derivation with respect to the argument and the index $0$ means that the function is evaluated at $r_0=R_0$.

Outside the star, $T_{\mu\nu}=0$ but the projected Weyl tensor is not necessarily zero. In keeping with the symmetry and staticity of the problem we write its components as
$${\cal E}^t_t=-\epsilon^e(r)\quad,\quad {\cal E}^r_r=\pi^e(r)\quad,\quad {\cal E}^\theta_\theta={\cal E}^\phi_\phi=\sigma^e(r)\quad,\quad r\geq r_0\,.$$
Similarly we write the stress-energy tensor of the star and the projected Weyl tensor in the star as
$$T^T_T=-\rho(R)\quad,\quad T^R_R=p(R)\quad,\quad T^\theta_\theta=T^\phi_\phi=s(R)\quad,\quad R\leq R_0$$
$${\cal E}^T_T=-\epsilon^i(R)\quad,\quad {\cal E}^R_R=\pi^i(R)\quad,\quad {\cal E}^\theta_\theta={\cal E}^\phi_\phi=\sigma^i(R)\quad,\quad R\leq R_0\,.$$
The continuity condition of the $(za)$ and $(zz)$ components of the right hand side of equation (1.1) ($z$ standing here for $r$ or $R$) is redundant with (3.1-3) but is useful and
reads as a condition on the matter on $S$
$$-\pi^e_0=8\pi Gp_0-\pi^i_0+{\kappa^2\over12}(\rho_0+s_0+p_0)(\rho_0+s_0-p_0)\,.\eqno(3.4)$$
In [5-7] a perfect fluid is assumed in the star (i.e. $p(R)=s(R)$) and the additional condition $p_0=0$ is imposed, yielding the conclusion that $\pi^e_0-\pi^i_0\neq0$, see [5]
and equation (3.4). In this paper we shall not impose $p_0$ to be necessarily zero.

In order to build a particular model, assumptions must be made about the equation of state of matter inside the star, and also about the projected Weyl tensor, which, in the
absence of information about the bulk, can be almost anything we like. If, for example ${\cal E}_{\mu\nu}$ is zero, then equation (1.1) reduces to Einstein's outside the star and
Birkhoff's theorem yields the Schwarzschild solution~: $\Phi^e(r)=\Psi^e(r)=1-2GM/r$ with $M$ the mass of the star. Another possibility, studied in [7-8] is that the metric
outside the star be of the Reisner-Nordstr\"om type~: $\Phi^e(r)=\Psi^e(r)=1-2GM/r+q/r^2$, with $q$ a constant not necessarily positive a priori. However, as we shall see in
Section V, if the metric in the bulk is to be asymptotically anti-de Sitter and regular, then the metric in the brane outside the star must tend, in the weak gravity regime
and for distances large compared to the bulk scale,  to the Randall-Sundrum solution [2] (see also [9-10]), that is (in  Schwarzschild coordinates)
$$\Phi^e(r)\simeq1-{2GM\over r}\left(1+{2\over3}{{\cal L}^2\over r^2}\right)\quad,\quad \Psi^e(r)\simeq1-{2GM\over r}\left(1+{{\cal L}^2\over r^2}\right)\eqno(3.5)$$
with ${\cal L}\equiv{\kappa\over8\pi G}$, and for $r\gg {\cal L}\,\,\, \hbox{and}\,\,\, {GM\over r}\ll 1$. Such a solution corresponds to the following (anisotropic) projected Weyl
tensor
$$\epsilon^e\simeq4GM{{\cal L}^2\over r^5}\quad,\quad\pi^e\simeq-2GM{{\cal L}^2\over r^5}\quad,\quad\sigma^e\simeq3GM{{\cal L}^2\over r^5}\,.\eqno(3.5bis)$$
The shortcomings of (3.5) are manifold~: first, it is an approximate solution, and the exact vacuum solution it is an approximation of is still unknown and probably not obtainable
in an analytical form~; second, it does not tell us what the projected Weyl tensor is like inside the star.

Proceeding nonetheless the gravity equations (1.1) read, inside the star and for matter being a perfect fluid ($p(R)=s(R)$) (cf [5] where they are written in a slightly different
form)
$$G{dm\over dR}=R^2\left(4\pi G\rho+{\kappa^2\over24}\rho^2-{1\over2}\epsilon^i\right)\eqno(3.6)$$
$$ R(R-2Gm){d\nu\over dR}=2Gm+R^3\left[8\pi G p-\pi^i+{\kappa^2\over12}\rho(\rho+2p)\right]\eqno(3.7)$$
$$(\rho+p){d\nu\over dR}=-2{dp\over dR}\eqno(3.8)$$
$${d\pi^i\over dR}+{1\over2}{d\nu\over dR}(\pi^i+\epsilon^i)+{2\over R}(\pi^i-\sigma^i)={\kappa^2\over6}{d\rho\over dR}(\rho+p)\eqno(3.9)$$
$$-\epsilon^i+\pi^i+2\sigma^i=0\eqno(3.10)$$
where we have set $\Psi^i(R)\equiv1- {2Gm(R)\over R}$ and $\Phi^i(R)\equiv e^{\nu(R)}$. 

For the sake of the example, consider the case of a constant density star and projected Weyl tensor given by
$$\rho(R)=\mu\quad,\quad p(R)=s(R)\quad\hbox{and}\quad\pi^i(R)=-{\alpha\over2}\epsilon^i(R)\quad,\quad\sigma^i(R)={\alpha+2\over4}\epsilon^i(R)\eqno(3.11)$$
where $\mu$ is a constant and $\alpha$ a parameter ($\alpha=1$ corresponds to a projected Weyl tensor having a similar form outside and inside the star). It must be clear
though that this example has no reason to be more realistic than those considered in [5] as there is no guarantee whatsoever that our choice for the projected Weyl tensor inside
the star yields a  metric in the bulk which is regular and asymptotically anti-de Sitter. Still proceeding however,  equations (3.8-9) integrate as
$$p=Ae^{-\nu/2}-\mu\quad,\quad \epsilon^i={\epsilon\over R^{3\alpha+2\over\alpha}}\,e^{-{(\alpha-2)\over2\alpha}\nu}$$
where $A$ and $\epsilon$ are integration constants. If we impose $\nu(R=0)=0$ (which we can always do) and $\epsilon^i(R=0)$ to be regular then $\alpha$ is restricted to be
in the range
$0>\alpha\geq-2/3$. To be specific we shall further restrict our attention to the isotropic case
$$\alpha=-{2\over3}\quad,\quad \epsilon^i=\epsilon\, e^{-2\nu}\quad,\quad \pi^i=\sigma^i={1\over3}\epsilon^i\,.$$
Equations (3.6-7) can then be integrated numerically or analytically as a power series. Indeed,
inserting the ansatz
$$\Phi^i(R)\equiv e^{\nu(R)}=1+a_1R+{1\over2}a_2R^2+...$$
in (3.6-7) we get
$$a_1=0\qquad,\qquad a_2=8\pi G\left(A-{2\over3}\mu\right)+{\kappa^2\over6}\mu\left(A-{1\over3}\mu\right)-{2\over3}\epsilon $$
and
$$\Psi^i(R)\equiv1-{2Gm(R)\over R}\qquad\hbox{with}\qquad Gm(R)={R^3\over3}\left(4\pi
G\mu+{\kappa^2\over24}\mu^2-{1\over2}\epsilon \right)+{a_2\over10}\epsilon R^5+...$$
as well as
$$p(R)=A-\mu-{1\over4}Aa_2R^2+...\qquad,\qquad\epsilon^i(R)=\epsilon (1-a_2R^2+...)\,.$$

 The junction conditions (3.1-4) then yield
the mass $M$ of the star in terms of its radius $r_0$, its density $\mu$ and its Weyl parameter $\epsilon $ as
$$M\simeq{4\pi \over3}\mu\left(1+{\kappa^2\over96\pi G}\mu-{\epsilon \over8\pi G\mu}\right)r^3_0\left(1-{{\cal L}^2\over r_0^2}\right)$$
and the surface pressure $p_0$ in terms of $\mu$ and $\epsilon^i_0\equiv \epsilon^i(r_0)$ as
$$2M{{\cal L}^2\over r_0^5}\simeq8\pi Gp_0-{1\over3}\epsilon^i_0+{\kappa^2\over12}\mu(\mu+2p_0)\,.$$
The surface pressure being thus known (and not necessarily zero) the integration constant $A$ is  determined and the pressure known everywhere. We shall not dwell on this
solution, especially since it is valid only in the weak gravity regime.
\bigskip

In the simpler (but even less realistic) case when the star density is constant and ${\cal E}_{\mu\nu}=0$ everywhere (so that the solution outside the star is Schwarzschild's)
then equations (3.6-10) integrate exactly (as in standard general relativity) as 
$$p=Ae^{-\nu/2}-\mu$$
with $A$ an integration constant and
$$ \Psi^i(R)\equiv1-{2Gm(R)\over R}\qquad\hbox{with}\qquad m(R)={4\pi \over3}\mu_{eff}R^3 \quad\hbox{and}\quad \mu_{eff}=\mu\left(1+{\kappa^2\mu\over96\pi
G}\right)$$
$$\Phi^i(R)\equiv e^{\nu(R)}=\left[ {3A_{eff}\over2\mu_{eff}}-D\sqrt{1-{8\pi G\over3}\mu_{eff}R^2}\right]^2\quad\hbox{with}\quad
A_{eff}=A\left(1+{\kappa^2\mu\over48\pi G}\right)$$
where $D$ can be chosen at will (e.g. $D=-1+3A_{eff}/2\mu_{eff}$).
Joigning this solution to Schwarzschild's (according to (3.1-4)) yields the mass $M$ of the star in terms of its radius $r_0$ and its effective density $\mu_{eff}$ as
$$M={4\pi\over3}\mu_{eff}r_0^3$$
and the surface pressure $p_0$ in terms of $\mu$ as
$$0=8\pi Gp_0+{\kappa^2\mu\over12}(\mu+2p_0)\,.$$
(Hence $p_0<0$...)  The integration constant $A$ is then determined as
$${A_{eff}\over2 D}=\mu_{eff}\sqrt{1-{2GM\over r_0}}$$
and the pressure is known everywhere as
$$p(R)=\mu\,{1+{\kappa^2\mu\over96\pi G}\over1+{\kappa^2\mu\over48\pi G}}{2\sqrt{1-{2GM\over r_0}}\over\left(3\sqrt{1-{2GM\over r_0}}-\sqrt{1-{2GMR^2\over
r_0^3}}\right)}-\mu\,.$$
The critical mass of the star (corresponding to an infinite central pressure) is given by $GM/r_0=4/9$ (as in standard general relativity, apart from the fact the $\mu$ is
replaced by $\mu_{eff}$).
\bigskip

In conclusion, the equations governing gravity in  a brane star are (3.6-10). They can be integrated once, as usual, an equation of state for the fluid is given (such as
$\rho=Const.$), and once assumptions (such as (3.11) or ${\cal E}_{\mu\nu}=0$) are made about the projected Weyl tensor. However, as long as some restrictions coming from
regularity conditions in the bulk are not  imposed (see Section V), the various models which can be built (such as those presented here) remain of limited physical interest.

\section{Collapsing ``stars" on a brane}

Let us now consider a spherically symmetric collapse situation. Outside the star, unless the projected Weyl tensor happens to be zero, in which case Birkhoff's theorem tells us
the solution must be Schwarzschild's, there is no reason a priori that the solution be even static, as emphasized in [6-7]. However we shall restrict here our attention, as in [6],
to the  (probably very) particular case when it is, and write the metric outside the star as before, that is as
$$ds^2_e=-\Phi(r)dt^2+{1\over\Psi(r)}dr^2+r^2d\Omega^2\,.$$ 
Should the occasion arise, we shall impose this metric to tend in the weak gravity regime to the Randall Sundrum solution (3.5). 

 The outer boundary $S_e$ of the collapsing star is defined parametrically as
$$r=r(\tau)\quad,\quad t=t(\tau)\,.$$
We choose the function $t(\tau)$ such that $\dot t=\sqrt{\Psi+\dot r^2}/\sqrt{\Phi\Psi}$, where a dot indicates
derivation with respect to
$\tau$, so that the induced metric on
$S_e$ reads
$$ds^2_e|_{S_e}=-d\tau^2+r^2(\tau)d\Omega^2$$ 
and the components of its extrinsic curvature are
$$^e\!K^\theta_\theta(\tau)=^e\!K^\phi_\phi(\tau)=-{1\over r}\sqrt{\Psi+\dot r^2}\quad,\quad ^e\!K^\tau_\tau(\tau)=-{1\over\sqrt{\Psi+\dot r^2}}\left[\ddot
r+{\Phi'\over\Phi}{\Psi\over2}+ {\dot r^2\over2}\left({\Phi'\over\Phi}-{\Psi'\over\Psi}\right)\right]$$
where a prime indicates a derivation with respect to $r$ and where $r=r(\tau)$.

We describe the interior of the collapsing star {\it \`a la} Oppenheimer-Snyder, that is by a homogeneous and isotropic matter distribution
so that the metric is of the Friedmann type
$$ds^2_i=-dT^2+a^2(T)(dR^2+R^2d\Omega^2)\,.\eqno(4.0)$$
(We take flat spatial sections for simplicity.)
The stress-energy tensor of matter inside the collapsing star depends on $T$ only and is conserved. Setting $T^T_T\equiv-\rho$ and
$T^R_R=T^\theta_\theta=T^\phi_\phi\equiv p\equiv w\rho$ we hence have
$${d\rho\over dT}+{3\over a}{d a\over dT}(1+w)\rho=0\,.\eqno(4.1)$$
In the particular case at hand the projected Weyl tensor is also conserved and, being traceless, behaves like a radiation fluid 
$$-{\cal E}^T_T\equiv\epsilon^i=-{3c\over a^4}\quad,\quad{\cal E}^R_R={\cal
E}^\theta_\theta={\cal E}^\phi_\phi\equiv\pi^i=-{c\over a^4}$$ with
$c$ a constant. As for the scale factor it satisfies the BDL  equation
(see [11]) (that is the gravity equation (1.1))
$${1\over a^2}\left({da\over dT}\right)^2={8\pi G\over3}\rho+{\kappa^2\rho^2\over36}+{c\over a^4}\,.\eqno(4.2)$$
Given an equation of state, that is, say, a numerical value for $w$, and given a numerical value for $c$, equations (4.1-2) gives $a(T)$ and $\rho(T)$. 

 The inner boundary $S_i$ of the collapsing star is defined parametrically as
$$R=R(\tau)\quad,\quad T=T(\tau)\,.$$
The function $T(\tau)$ is chosen such that $\dot T=\sqrt{1+a^2\dot R^2}$ so that the induced metric on $S_i$ reads
$$ds^2_i|_{S_i}=-d\tau^2+a^2R^2d\Omega^2$$ 
and the components of its extrinsic curvature are
$$^i\!K^\theta_\theta(\tau)=^i\!K^\phi_\phi(\tau)=\dot R{da\over dT}+{\sqrt{1+a^2\dot R^2}\over Ra}\quad,\quad ^i\!K^\tau_\tau(\tau)={a\ddot R\over\sqrt{1+a^2\dot R^2}}
+2\dot R{da\over dT}$$
where  $R=R(\tau)$ and $a=a(T(\tau))$.

Now the junction conditions, that is the continuity of the induced metrics and extrinsic curvatures of the boundary of the star,
read
$$r=aR\eqno(4.3)$$
$$(aR)^2\left({1\over a}{da\over dT}\right)^2=1-\Psi(aR)\eqno(4.4)$$
$$\dot r\sqrt{\Psi+\dot r^2}\left({\Psi'\over\Psi}-{\Phi'\over\Phi}\right)=2R\dot R\dot T\left[a{d^2a\over dT^2}-\left({da\over dT}\right)^2\right]\,.\eqno(4.5)$$
In standard general relativity, the metric outside the star is necessarily Schwarzschild's~: $\Phi=\Psi=1-2GM/r$ so that (4.5) imposes $R(\tau)$ to be a constant. Equation (4.4)
then says that $a(T)\propto T^{2/3}$ and the standard Friedmann equation $\left[{1\over a^2}\left({da\over dT}\right)^2={8\pi G\over3}\rho\right]$ together with the
conservation equation (4.1) then yields the well-known result that the matter inside the collapsing star must be
dust~: $w=0$.

In the case of a star on  a brane on the other hand, equation (4.5) tells us that $R(\tau)$ is constant only if $\Phi=\Psi$ which is the case if the metric outside the star is chosen
to be Schwarzschild's or Reisner-Nordstr\"om's. Then equation (4.4) gives us the time dependence of the scale factor $a(T)$, for example $a\propto T^{2/3}$ if the exterior is
Schwarzschild's. The BDL equation (4.2) then gives us the time dependence of the matter density $\rho$ and the conservation equation (4.1) yields the equation of state, that is
$w(T)$, which is {\it not} zero, not even a constant (and behaves strangely~; for example $w(a=0)<0$ if $c\leq0$).

If now one imposes the solution outside the star to be the Randall-Sundrum solution (3.5) then, as emphasized in [6-7], $\dot R\neq0$ and matter inside the star cannot be dust.
However this does not necessarily imply that the solution outside the star cannot be static, as argued in [6-7]. Indeed equations (4.3-5) do have a solution~: they  are coupled
equations for $R(\tau)$ and $a(T)$ and the BDL equations (4.1-2) then give $\rho(T)$ and the equation of state $w(T)$.

In conclusion, if the collapsing star is described by the BDL equation and {\it if} the exterior of the collapsing star is static then the matching conditions (4.3-5) give the
motion of the star boundary $R(\tau)$ (which is not, in general, in free fall) as well as the equation of state of the matter in the star (which is not dust, contrarily to the
standard general relativistic Oppenheimer-Snyder model). Now, again, the behaviour of the bulk must be analyzed and checked to be  asymptotically anti-de Sitter and non
singular before the models can be given serious physical content.

\section{The view from the bulk~: a toy model}

As we have already amply emphasized, gravity on the brane depends on gravity in the bulk via the projected Weyl tensor. Now, the equations for gravity in the bulk, as well as
their allowed solutions, should spring from the (yet to be built) ``Grand Theory" underlying the five dimensional effective brane world picture studied here. To be specific, and in
accordance with common views (see e.g. [2] and references therein), we shall impose that the bulk is an Einstein space, is asymptotically anti-de Sitter and is free of curvature
singularities (at least outside the brane).

We considered in this paper a bulk whose boundary (the brane) is divided into two regions, the inside and the outside of a spherically symmetric star. The bulk is hence itself
divided into two regions, one, ${\cal M}_e$, which ``projects" onto the brane on the outside region of the star, the other, ${\cal M}_i$ which projects on the inside of the star, the
two regions being delineated by a kind of ``tube", $\Sigma$, which projects on the boundary $S$ of the star.

As we have seen, many solutions are possible on the brane outside the star, depending on the assumptions made about the projected Weyl tensor ${\cal E}_{\mu\nu}$. For
example, if we suppose that ${\cal E}_{\mu\nu}=0$, then the solution on the brane outside the star is Schwarzschild's and, as shown by Chamblin et al. in [12], the metric in the
Einstein bulk
${\cal M}_e$ outside the tube $\Sigma$ can be written, in gaussian normal coordinates $(y,t,r,\theta,\phi)$ where the equation for the brane is $y=0$, as
$$d\sigma^2_e=dy^2+e^{-2 y/{\cal L}}\left[-\left(1-{2GM\over r}\right)dt^2+{1\over\left(1-{2GM\over r}\right)}dr^2+r^2d\Omega^2\right]\,.\eqno(5.1)$$
Now this solution does not fulfill the imposed criteria since it exhibits a curvature singularity for all $r$ at $y\to\infty$
[12] and must be rejected on this ground. One should rather choose the vacuum, spherically symmetric and static Randall-Sundrum solution [2] (which projects as (3.5) on the brane
outside the star and whose explicit expression in the bulk can be found
in, e.g.,  [9-10]) since it  is almost anti-de Sitter everywhere in the bulk ${\cal M}_e$ and hence is an acceptable
solution. Unfortunately the static exact solution it is an approximation of is yet unknown. We do not even know if this yet to be found static exact solution is unique and
everywhere regular (note however that numerical calculations probing these points are underway, Turok and Wiseman, private communication). Finally, even less is known about
bulk solutions which would project on non static solutions outside the star, such as the Vaidya metric considered in [7]...

Concerning now the solution inside the star on the brane and inside the tube $\Sigma$ in the bulk ${\cal M}_i$, the situation is just as bleak. In the static case the non-vacuum,
spherically symmetric, almost anti-de Sitter Randall-Sundrum solution [2] [9] can be obtained in terms of the Fourier transform of $T_{\mu\nu}$ but is quite
awkward to handle (especially since the Fourier transform of the Heaviside distribution is involved). We are better off in the case of a collapsing homogeneous
and isotropic star described by equations (4.1-2) since the corresponding bulk solution is the BDL metric [11], whose (complicated) explicit expression in gaussian normal
coordinates $(y,T,R,\theta,\phi)$ can be found in [11] but which turns out to be nothing but the five dimensional anti-de Sitter solution if $c=0$ and the Schwarzschild anti-de
Sitter solution if $c\neq0$ (see e.g. [13]).

Assuming we know the bulk metrics, that is assuming for example that ${\cal M}_e$ is the Randall-Sundrum solution  (or the Chamblin et al. exact solution (5.1) if we relax
our criteria), and assuming that
${\cal M}_i$ is, say, a (Schwarzschild) anti-de Sitter spacetime, the question then becomes~: can one join these two bulks along their common boundary
$\Sigma$ without introducing a thin layer of matter~? In other words, can we find  boundaries $\Sigma_e$ and $\Sigma_i$ to ${\cal M}_e$  and ${\cal M}_i$ whose induced
metrics and extrinsic curvatures are the same~?
\bigskip

We leave this question for another work and shall content ourselves here with the following three dimensional toy model which, we hope, sheds some light on how to proceed.

Let us consider a three dimensional ``bulk" ${\cal M}_3$ (with coordinates $x^A$) which satisfies, everywhere but on a ``brane" $M_2$, some ``Einstein equations", say ${\cal
R}=0$ where ${\cal R}$ is the scalar curvature of ${\cal M}_3$. An example of such a bulk is
$${\cal M}_e\quad\hbox{with metric}\quad d\sigma^2_e=dz^2+a^2z^2dr^2+r^2d\phi^2\eqno(5.2)$$
with $a$ a constant. This bulk is non flat and even singular since ${\cal R}_{ABCD}{\cal R}^{ABCD}=(arz^2)^{-2}$. It will play in our toy model the role, say, of the Chamblin et al.
solution (5.1). An another example is $${\cal M}_i\quad\hbox{with metric}\quad d\sigma^2_i=dZ^2+d\rho^2+\rho^2d\phi^2\eqno(5.3)$$
which is flat and will play the role, say, of the five dimensional anti-de Sitter spacetime. Performing the change of coordinates
$\rho\to\tilde\rho$, $Z\to\tilde Z$ such that 
$$\rho={\tilde\rho-Z'_i(\tilde\rho)\tilde Z\over\sqrt{1+Z^{'2}_i(\tilde\rho)}}\quad,\quad Z=Z_i(\tilde\rho)+{\tilde Z\over\sqrt{1+Z^{'2}_i(\tilde\rho)}}\eqno(5.4)$$
where  $Z_i(\tilde\rho)$ is an arbitrary function, the flat metric (5.3) takes the form
$$d\sigma^2_i=d\tilde Z^2+\left(1+Z^{'2}_i\right)\left[1-{Z''_i\tilde Z\over\left(1+Z^{'2}_i\right)^{3/2}}\right]^2d\tilde\rho^2+\left[\tilde\rho-{Z'_i\tilde
Z\over\sqrt{1+Z^{'2}_i}}\right]^2d\phi^2\eqno(5.5)$$
 which we shall interpret as the toy analogue of the BDL metric.

Now, the $(2+1)$ decomposition of ${\cal R}=0$ yields the equation for ``gravity" in the ``brane" $M_2$ (coordinates $x^\mu$) as
$$R=K^2+K\,.\,K-2n^A\partial_A{\cal K}|_{M_2}\eqno(5.6)$$
where $R$ is the scalar curvature of $M_2$, where $K_{\mu\nu}$ is its extrinsic curvature, which is interpreted as being related to the ``stress-energy tensor of matter" in the
brane, and where
$n^A\partial_A{\cal K}|_{M_2}$ is the ``projected Weyl tensor". This equation is the toy analogue of equation (1.1). An example of such a brane $M_e$ is the surface
$z=Const.=z_e$ in ${\cal M}_e$, with metric
$$d\sigma^2_e|_{M_e}\equiv ds^2_e=a^2z_e^2dr^2+r^2d\phi^2\,.\eqno(5.7)$$
$M_e$ is locally flat. It is a cone with extrinsic curvature
$$^e\!K_{\phi\phi}=0\quad,\quad ^e\!K_{rr}=-a^2z_e\,.\eqno(5.8)$$
This non zero extrinsic curvature is interpreted in this toy model as some ``tension", and the brane $M_e$ as a ``vacuum solution" of the brane gravity equation (5.6). In other
words (5.8) will be the toy equivalent of the Schwarszchild solution. Another example of a brane $M_i$ is the surface $Z=Z_i(\rho)$, or equivalently
$\tilde Z=0$, in
${\cal M}_i$. The induced metric on $M_i$ and its extrinsic curvature are
$$d\sigma^2_i|_{M_i}\equiv ds^2_i=(1+Z^{'2}_i)d\rho^2+\rho^2d\phi^2\eqno(5.9)$$
$$^i\!K_{\phi\phi}={\rho Z'_i\over\sqrt{1+Z^{'2}_i}}\quad,\quad ^i\!K_{\rho\rho}={Z''_i\over\sqrt{1+Z^{'2}_i}}\,.\eqno(5.10)$$
$M_i$ is our toy equivalent of the Friedmann metric inside the star, $Z_i(\rho)$ playing the role of the scale factor.
The extrinsic curvature being interpreted as the stress-energy tensor of matter in the brane, one now chooses some ``equation of state", that is a relation between
$^i\!K_{\phi\phi}$ and $ ^i\!K_{\rho\rho}$ which then determines $Z_i(\rho)$. For example, if one imposes, say $^i\!K^\phi_\phi= ^i\!K^\rho_\rho$, then we have
$$Z_i(\rho)=A-\sqrt{B-\rho^2}\quad\hbox{and}\quad ds^2_i={B\over B-\rho^2}d\rho^2+\rho^2d\phi^2\eqno(5.11)$$
$A$ and $B$ being constants. Such a brane metric is to be seen as playing the role of, say, the dust brane BDL solution.

Let us now describe the ``star" that is the circles $S_e$ and $S_i$ which are the boundaries of $M_e$ and $M_i$.

In $M_e$ the star is the circle $r=r_0$ with induced metric $ds^2_e|_{S_e}=r^2_0d\phi^2$ and extrinsic curvature $^e\!k_{\phi\phi}=-r_0/az_e$. In $M_i$ the star is the circle
$\rho=\rho_0$ with induced metric $ds^2_i|_{S_i}=\rho^2_0d\phi^2$ and extrinsic curvature $^i\!k_{\phi\phi}=-\rho_0/\sqrt{1+Z^{'2}_i|_0}$. Repeting the arguments of Section II
we shall assume that the junction conditions between the exterior and the interior of the star are that the induced metrics and extrinsic curvatures of $S_e$ and $S_i$ be the
same. Hence we must have
$$r_0=\rho_0\quad,\quad a^2z_e^2=1+Z^{'2}_i|_0\,.\eqno(5.12)$$
In our toy model, $az_e$ is the analogous of the Schwarszchild mass and $Z_i$ the analogous of a scale factor. Hence (5.12) is similar to the junction conditions which, in
standard general relativity, tell us that matter inside the star must be dust and relate the Schwarszchild mass to the radius of the star.

What remains to be done is to join the bulks ${\cal M}_e$ and ${\cal M}_i$.

To do so, consider the family of surfaces $\Sigma_e$ in ${\cal M}_e$ defined by $z=\zeta_e(r)$. The induced metric on $\Sigma_e$ and its intrinsic curvature are
$$d\sigma^2_e|_{\Sigma_e}=(a^2\zeta_e^2+\zeta^{'2}_e)dr^2+r^2d\phi^2\eqno(5.13)$$
$$^e\!{\cal K}_{\phi\phi}=-{r\zeta'_e\over a\zeta_e\sqrt{a^2\zeta^2_e+\zeta^{'2}_e}}\quad,\quad ^e\!{\cal K}_{rr}=-{a\over\sqrt{a^2\zeta^2_e+\zeta^{'2}_e}}
(a^2\zeta_e^2+2\zeta^{'2}_e-\zeta_e\zeta''_e)\eqno(5.14)$$
(and the boundary $S_e$ of the brane star is the circle $r=r_0$ in $\Sigma_e$.)
Consider also the family of surfaces $\Sigma_i$ in ${\cal M}_i$ defined by $Z=\zeta_i(\rho)$. The induced metric on $\Sigma_i$ and its intrinsic curvature are
$$d\sigma^2_i|_{\Sigma_i}=(1+\zeta^{'2}_i)d\rho^2+\rho^2d\phi^2\eqno(5.15)$$
$$^i\!{\cal K}_{\phi\phi}={\rho\zeta'_i\over\sqrt{1+\zeta^{'2}_i}}\quad,\quad ^i\!{\cal
K}_{\rho\rho}={\zeta''_i\over\sqrt{1+\zeta^{'2}_i}}\eqno(5.16)$$
(and the boundary $S_i$ of the brane star is the circle $\rho=\rho_0$ in $\Sigma_i$.) 

According to our criteria the boundary between ${\cal M}_e$ and ${\cal M}_i$ must be regular, that is the induced metrics and extrinsic curvatures of $\Sigma_e$ and
$\Sigma_i$ must be the same. Comparing (5.13-14) to (5.15-16) this imposes $\zeta_e=1/a$ and $\zeta_i=Const.$, that is it picks up the surfaces $\Sigma_e$ and
$\Sigma_i$ to be two dimensional planes. Hence our toy problem has a  solution~: ${\cal M}_e$ and ${\cal M}_i$ can be smoothly joined.

\bigskip
The lesson to draw from this (over simplified~!) model when treating stars on branes is that probably the main restriction comes from finding regular and asymptotically anti-de
Sitter Einstein bulks which project onto the exterior and the interior of the star, and not from smoothly joining these bulks, as it stems from our toy model that one can
find a smooth boundary between them.

\section{Acknowledgements}

I warmly thank Joseph Katz with whom I discussed most aspects of this work. I am grateful to Daisuke Ida, Keichi Maeda, Tetsuye Shiromizu, and Takahiro Tanaka for discussions
about brane world in general, and  to Roy Maartens, Tsvi Piran, Neil Turok and Tobby Wiseman for discussions about brane stars in particular. I also warmly thank
Elisabeth Jacobs for her hospitality in Villepreux where  this work was completed.

\end{document}